\documentstyle{article}

\newcommand{\be}{\begin{equation}}
\newcommand{\ee}{\end{equation}}
\newcommand{\bea}{\begin{eqnarray}}
\newcommand{\eea}{\end{eqnarray}}
\newcommand{\beas}{\begin{eqnarray*}}
\newcommand{\eeas}{\end{eqnarray*}}
\newcommand{\bdm}{\begin{displaymath}}
\newcommand{\edm}{\end{displaymath}}
\newcommand{\ba}{\begin{array}}
\newcommand{\ea}{\end{array}}
\newcommand{\bi}{\begin{itemize}}
\newcommand{\ei}{\end{itemize}}
\newcommand{\ben}{\begin{enumerate}}
\newcommand{\een}{\end{enumerate}}
\newcommand{\bc}{\begin{center}}
\newcommand{\ec}{\end{center}}
\newcommand{\bfl}{\begin{flushleft}}
\newcommand{\efl}{\end{flushleft}}
\newcommand{\bfr}{\begin{flushright}}
\newcommand{\efr}{\end{flushright}}
\newcommand{\bd}{\begin{description}}
\newcommand{\ed}{\end{description}}
\newcommand{\bq}{\begin{quote}}
\newcommand{\eq}{\end{quote}}
\newcommand{\bfg}{\begin{figure}}
\newcommand{\efg}{\end{figure}}
\newcommand{\bt}{\begin{table}}
\newcommand{\et}{\end{table}}
\newcommand{\btb}{\begin{tabular}}
\newcommand{\etb}{\end{tabular}}
\newcommand{\btg}{\begin{tabbing}}
\newcommand{\etg}{\end{tabbing}}

\begin{document}
\title{\raisebox{2cm}{\makebox[0pt][r]{\small UTAS-PHYS-94-23}}
A Short Note on Two-loop Box Functions}
\author{Dirk Kreimer\thanks{email: kreimer@physvax.phys.utas.edu.au}\\
\small Dept.~of Physics\\
\small Univ.~of Tasmania\\
\small G.P.O.Box 252C\\
\small Hobart, 7001\\
\small Australia}
\date{June 1994}
\maketitle
\begin{abstract}
It is shown that the two-loop four-point functions are similar in structure
to the three-point two-loop functions for all mass cases and topologies.
The result is derived by using a rotation to a $(+,-,-,+)$ signature
without spoiling analyticity properties.
\end{abstract}

In this short note we will discuss some similarities between two-loop
three- and four-point functions. It was first noticed in \cite{davydychev}
that there is a striking similarity between the special functions
appearing in the three- and four-point cases. But this result was derived
only for the very restricted mass case of vanishing internal masses.
In this note we will
show that the two-loop box functions allow for integral representations
similar to the ones obtained for the three-point case in all mass cases
and for all possible topologies. Following the lines of \cite{CKK}
we hope to be able to come up with explicit numerical results for
arbitrary two-loop box functions in the future \cite{CK}.

The paper is organized as follows. We first show how to transform
the integrals via an unconventional Wick rotation to a $(+,-,-,+)$
signature.
We explicitly show that this will not spoil any analyticity properties of
the Green functions. In the next step, we show that the three-point and
four-point function (and also the two-point function) can be treated on
the same footing. This is based on the
fact that both cases allow for similar two- and three-particle cuts.
The results turn out to be three-fold integral representations similar to
the ones obtained in \cite{dirk3p} for the three-point functions.
These can then be transformed into convenient integral representations
along the lines of \cite{CKK}.

All possible two-loop functions can be derived from the following topology
Fig.(1) by assigning two, three, or four external particles in all possible
ways
to it.\\[5mm]
%fig1
%_________________________________________________________________________
% FIGURE
%
\unitlength=1.00mm
%\special{em:linewidth 0.4pt}
\linethickness{0.4pt}
\begin{picture}(33.00,14.00)
\put(16.00,7.00){\circle{14.00}}
\put(16.00,13.00){\line(0,-1){13.00}}
\put(16.00,13.00){\line(0,1){1.00}}
\end{picture}\nopagebreak
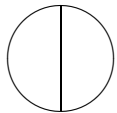
{\small Fig.1: The generic two-loop
topology.\\[5mm]}

Consider for example the four-point functions. We have four
different topologies:\\[5mm]
%fig2
%_________________________________________________________________________
% FIGURE
%
\unitlength=1.00mm
%\special{em:linewidth 0.4pt}
\linethickness{0.4pt}
\begin{picture}(133.00,14.00)
\put(16.00,7.00){\circle{14.00}}
\put(16.00,13.00){\line(0,-1){13.00}}
\put(16.00,13.00){\line(0,1){1.00}}
\put(9.50,8.00){\line(-1,0){4.00}}
\put(9.50,6.00){\line(-1,0){4.00}}
\put(22.50,8.00){\line(1,0){4.00}}
\put(22.50,6.00){\line(1,0){4.00}}
\put(46.00,7.00){\circle{14.00}}
\put(46.00,13.00){\line(0,-1){13.00}}
\put(46.00,13.00){\line(0,1){1.00}}
\put(39.50,8.00){\line(-1,0){4.00}}
\put(39.50,6.00){\line(-1,0){4.00}}
\put(52.50,8.00){\line(1,0){4.00}}
\put(46.00,7.00){\line(1,0){4.00}}
\put(76.00,7.00){\circle{14.00}}
\put(76.00,13.00){\line(0,-1){13.00}}
\put(76.00,13.00){\line(0,1){1.00}}
\put(70.00,9.00){\line(-1,0){4.00}}
\put(70.00,5.00){\line(-1,0){4.00}}
\put(69.00,7.00){\line(-1,0){4.00}}
\put(83.00,7.00){\line(1,0){4.00}}
\put(106.00,7.00){\circle{14.00}}
\put(106.00,13.00){\line(0,-1){13.00}}
\put(106.00,13.00){\line(0,1){1.00}}
\put(100.00,9.00){\line(-1,0){4.00}}
\put(100.00,5.00){\line(-1,0){4.00}}
\put(99.50,8.00){\line(-1,0){4.00}}
\put(99.50,6.00){\line(-1,0){4.00}}
\end{picture}\\
%\vspace{5mm}
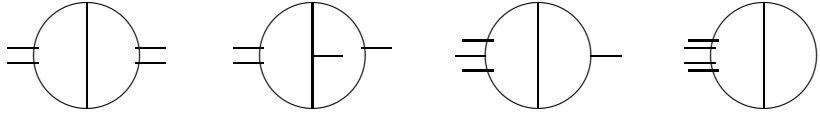
{\small Fig.2a,$\ldots$,d: The two-loop
four-point topologies.\\[5mm]}

They can all be written in the form (our only objective being scalar
integrals)
\bea
\int d^4 l \int d^4 k \frac{1}{N_l}\frac{1}{N_{l,k}}\frac{1}{N_k}
\label{eq1}
\eea
where
\begin{itemize}
\item
for Fig.(2a) we have:
\beas
N_l & = &  (l^2-m_1^2+i \eta)((l+p_1)^2-m_2^2+i\eta)((l+p_1+p_2)^2-m_3^2
+i\eta)\\
N_{l,k} & = & (l+k)^2-m_4^2+i\eta\\
N_k & = &  (k^2-m_5^2+i \eta)((k+p_3)^2-m_6^2+i\eta)((k-p_1-p_2)^2-m_7^2
+i\eta),
\eeas
\item
for Fig.(2b) we have:
\beas
N_l & = &  (l^2-m_1^2+i \eta)((l+p_1)^2-m_2^2+i\eta)((l+p_1+p_2)^2-m_3^2
+i\eta)\\
N_{l,k} & = & ((l+k)^2-m_4^2+i\eta)((l+k+p_3)^2-m_5^2+i\eta)\\
N_k & = &  ((k+p3)^2-m_6^2+i \eta)((k-p_1-p_2)^2-m_7^2
+i\eta),
\eeas
\item
for Fig.(2c) we have
\beas
N_l & = &  (l^2-m_1^2+i \eta)((l+p_1)^2-m_2^2+i\eta)((l+p_1+p_2)^2-m_3^2
+i\eta)\\
 & & ((l+p_1+p_2+p_3)^2-m_4^2+i\eta)\\
N_{l,k} & = & (l+k)^2-m_5^2+i\eta\\
N_k & = &  (k^2-m_6^2+i \eta)((k-p_1-p_2-p_3)^2-m_7^2
+i\eta),
\eeas
\item
while for Fig.(2d) we have
\beas
N_l & = &  (l^2-m_1^2+i \eta)((l+p_1)^2-m_2^2+i\eta)((l+p_1+p_2)^2-m_3^2
+i\eta)\\
 & & ((l+p_1+p_2+p_3)^2-m_4^2+i\eta)(l^2-m_5^2+i\eta)\\
N_{l,k} & = & (l+k)^2-m_6^2+i\eta\\
N_k & = &  (k^2-m_7^2+i \eta)
+i\eta).
\eeas
\end{itemize}
Here and in the following it is always understood that the integration
is split into an integration over the span of the exterior momenta (the
parallel space) and an integration over its orthogonal complement
(the orthogonal space) \cite{dirk2p,dirk3p}.
The propagators are quadratic forms in four resp.~eight variables each.
Note that
the fourth variable ($l_3$ resp.~$k_3$ which we can assume to
coincide with the orthogonal space variables) does not mix with components
of exterior momenta. As functions of $l_3,k_3$ our
denominators behave as
\beas
N_l & = & N_l(l_3^2)\\
N_{l,k} & = & N_{l,k}((l_3+k_3)^2)\\
N_k & = & N_k(k_3^2).
\eeas

We further stress that also the two-loop two- and three-point functions
fit into the template of Eq.(\ref{eq1}). If we consider the number of
propagators and the number of variables that do not mix with
exterior momenta
in the propagators, it is only in these numbers where the
two-, three- and four-point functions differ from our viewpoint.

Now let us do the following  transformations:
\beas
\int_{-\infty}^{+\infty}dl_3
\int_{-\infty}^{+\infty}dk_3
\frac{1}{N_l(l_3^2)N_{l,k}((l_3+k_3)^2)N_k(k_3^2)}\\
= 2\int_0^{+\infty}dl_3 \int_0^{+\infty}dk_3
\left(
\frac{1}{N_l(l_3^2)N_{l,k}((l_3+k_3)^2)N_k(k_3^2)}
\right. \\
\left.
+\frac{1}{N_l(l_3^2)N_{l,k}((l_3-k_3)^2)N_k(k_3^2)}
\right) \\
=\frac{1}{2}\int_0^{+\infty}\frac{ds}{\sqrt{s}}
\int_0^{+\infty}\frac{dt}{\sqrt{t}}
\left(
\frac{1}{N_l(s)N_{l,k}((\sqrt{s}+\sqrt{t})^2)N_k(t)}
\right. \\
\left.
+\frac{1}{N_l(s)N_{l,k}((\sqrt{s}-\sqrt{t})^2)N_k(t)}
\right) \\
=\int_0^1 du\int_0^{+\infty}v d\!v
\left(
\frac{1}{N_l(uv^2)N_{l,k}(v^2(\sqrt{u}+\sqrt{1-u})^2)N_k((1-u)v^2)}
\right. \\
\left.
+\frac{1}{N_l(uv^2)N_{l,k}(v^2(\sqrt{u}-\sqrt{1-u})^2)N_k((1-u)v^2)}
\right) .
\eeas
Using
that $(u,(1-u),(\sqrt{u} {+ \atop -} \sqrt{1-u})^2)$
all are $\geq 0$ between
zero and one we see that, for all propagators,
the poles in the $v$ variable
are located in the first or third quadrant (in the complex $v$-plane).
So we rotate $\pi/2$ clockwise in the $v$-plane and invert all the above
transformations
to obtain
\beas
\int_{-\infty}^{+\infty}dl_3
\int_{-\infty}^{+\infty}dk_3
\frac{1}{N_l(l_3^2)N_{l,k}((l_3+k_3)^2)N_k(k_3^2)}\\
=
\int_{-\infty}^{+\infty}dl_3
\int_{-\infty}^{+\infty}dk_3
\frac{1}{N_l(-l_3^2)N_{l,k}(-(l_3+k_3)^2)N_k(-k_3^2)}.
\eeas

We see that we effectively have transformed our Green functions from a
metric with signature $(+,-,-,-)$ to a signature $(+,-,-,+)$.
As this is quite an unusual transformation we have given the explicit
derivation above to show that this is possible without spoiling
analyticity properties. Note further
that this does not restrict the domain of validity of our results. This is
in contrast with the standard Wick rotation \cite{Itz}:

There one first considers an euclidean domain $q_i^2 <0$ for all exterior
momenta $q_i$. That guarantees that all these euclidean vectors
are orthogonal to some timelike vector $n$ say. Lorentz invariance of the
Green functions allows to choose $n=(1,0,0,0)$. In this frame the $0$-component
of all euclidean vectors $q_i$ vanishes, so that all propagators are quadratic
forms
where the $0$-variables ($l_0$ resp.~$k_0$) do not mix with exterior
momenta. This then guarantees that the causal behaviour of all propagators
is such that the poles of all propagators are located either in the second
or fourth quadrant of the complex $l_0$- resp.~$k_0$-plane.
Consequently, this allows
for a Wick rotation in these variables, which would otherwise be spoilt
by the 0-components of the exterior momenta. Indeed, these components would
shift
the poles into the other quadrants. So one ends up with the
result that for exterior euclidean momenta
one can Wick rotate to a metric of definite signature. One recovers then
the Green function for arbitrary timelike exterior momenta by an appropriate
analytic continuation. With this step one touches some fundamental results
of field theory, e.g.~the ability to recover the Wightman functions from
the Schwinger functions \cite{strocchi}.

In our case the situation is much simpler. Due to the fact that we can choose
a basis where the 3-variables $l_3$ and $k_3$ are free of exterior momenta
anyhow, we can directly do our transformation without posing
any restriction on the exterior momenta, overcoming the need of an
analytic continuation at the end altogether.

Nevertheless, we like to stress, as we are deforming to a still nondefinite
signature, we are not allowed to set the small imaginary parts of the
propagators to zero. They reflect the fact that the Green functions
of field theory are boundary values of analytic functions (where the
boundary is approached in the limit $\eta\rightarrow 0$ \cite{thiess,
wightman}). As we will make use of standard results of complex analysis in
further steps, we have to keep the $\eta$ prescription until the end of
our calculation to be allowed to handle the integrand as an analytic function.

Having this in mind, let us investigate the behaviour of two-, three-, and
four-point two-loop functions now. They differ only by the dimension of the
parallel space \cite{dirk2p,dirk3p,dirkt} and the number of propagators
contributing to $N_l,N_k,N_{l,k}$. With appropriate partial fractions
done in $N_l, N_k, N_{l,k}$ separately we can
reduce all cases to sums of terms of the following structure:
\beas
\int d^4l \int d^4 k\frac{1}{\Pi}\frac{1}{P_l P_{l,k}P_k},\\
\mbox{with} P_l\in N_l,\!P_{l,k}\in N_{l,k},\! P_{k}\in N_k.
\eeas
Here $\Pi$ is a function of parallel space variables only and is
a product of differences of propagators.
We can use translation invariance to bring the propagators
into a standard form:
\beas
P_l & = & l_0^2-l_1^2-l_2^2+l_3^2-m_l^2+i\eta\\
P_{l,k} & = & (l_0+k_0)^2-(l_1+k_1)^2-(l_2+k_2)^2+(l_3+k_3)^2-m_{l,k}^2
+i\eta\\
P_k & = & k_0^2-k_1^2-k_2^2+k_3^2-m_k^2+i\eta.\\
\eeas
Next we use translation invariance once more to do the following
shifts:
\beas
l_0\rightarrow l_0+l_1,\;l_3\rightarrow\;l_3+l_2,\\
k_0\rightarrow k_0+k_1,\;k_3\rightarrow\;k_3+k_2,
\eeas
which results in a representation of the form
\beas
\int d^4l \int d^4 k\frac{1}{\tilde{\Pi}}\frac{1}{
(l_0^2+2l_0l_1+2l_2l_3+l_3^2-m_l^2+i\eta)}\\
%% FOLLOWING LINE CANNOT BE BROKEN BEFORE 80 CHAR
\frac{1}{((l_0+k_0)^2+2(l_0+k_0)(l_1+k_1)+2(l_2+k_2)(l_3+k_3)+(l_3+k_3)^2-m_{l,k}^2
+i\eta)}\\
\frac{1}{(k_0^2+2k_0k_1+2k_2k_3+k_3^2-m_k^2+i\eta)},
\eeas
with some translated $\tilde{\Pi}$.

We recognize that the propagators are all linear in the middle
variables $l_1,l_2,k_1,k_2$. The coefficients depend solely on the
edge variables $l_0,l_3,k_0,k_3$. In particular the location of the
imaginary parts of the propagators, considered as functions of the
middle variables, depends on the sign of these coefficients.
That is, for positive $l_0,k_0,l_3,k_3$ resp.~$(l_0+k_0),(l_3+k_3)$
the middle variables $l_1,k_1,l_2,k_2$ have their poles in
the lower
half-plane and vice versa.

We now intend to do the integration in the middle variables
via
the residue theorem. We choose to close in the upper half-plane
for all four variables,
which imposes constraints on the domain of
integration for the edge variables \cite{dirk3p,CKK}.
This is so because only
for certain domains of the edge variables the poles of our propagators
will be located in the upper half-plane, as we have just seen.
We have to change the order of integration to do the integration in the four
middle variables $l_1,l_2,k_1,k_2$ first. We are allowed to do so
because the integral over the modulus of our integrand exists
\beas
\int d^4l \int d^4 k\frac{1}{\mid\tilde{\Pi}}\frac{1}{
\tilde{P_l}}
\frac{1}{\tilde{P_{l,k}}}
\frac{1}{\tilde{P_k}\mid}< \infty,
\eeas
as can be easily checked by explicit calculation. This is not necessarily
so (for example doing the same steps as above) in the one-loop case
$N_k=N_{l,k}=1$; one runs into trouble for the two- and three-point case.
There, the low dimensional parallel space results in a degenerate
$\tilde{\Pi}$ which does not depend on $l_2,l_3$ so that we cannot
exchange the $l_3,l_2$ integration. In the two-loop case this
degeneracy is cancelled by the mixed propagator $P_{l,k}$.
So in the case considered here
we can interchange the orders of integration and obtain a result as a sum
over terms of the form
\beas
\int_{d_{l_3}}^{u_{l_3}}
\int_{d_{k_3}}^{u_{k_3}}
\int_{d_{l_0}}^{u_{l_0}}
\int_{d_{k_0}}^{u_{k_0}}
d\!k_0d\!l_0d\!k_3d\!l_3
\frac{1}{Q(l_0,k_0,l_3,k_3)}
\eeas
where the upper and lower boundaries $(u_{\dots},d_{\ldots})$
are determined by the constraints
mentioned above
and the new function $Q$ is a quadratic form in the edge variables.

The actual form of $Q$, the boundaries and the number of contributing
terms are governed by the topology and number of external particles
coupling to the topology of Fig.(1) but the above structure is
generic to all two-loop functions. It is only the fact that the
function $\Pi$ can be kept clear of all middle variables in the case
of the two-point function which allows for such an easy integral representation
(integrating out the edge variables $l_3,k_3$) of this
master function \cite{dirk2p}. The factor
\beas
\frac{1}{(P_1-P_2)(P_4-P_5)}
\eeas
appearing there is just the function $\Pi$ depending
on the two edge-variables $l_0,k_0$ arising here.

Neither for the three- nor for the four-point function does
such a simplification
take place. Instead we have to integrate our quadratic forms $Q$
which gives us integral representations of the form
\beas
\int\int\int \frac{\log(Q_1+\sqrt{Q_2})}{\sqrt{Q_2}Q_3},
\eeas
with quadratic forms $Q_i$
for both cases. Integral representations of this kind are familiar for
the three-point case \cite{dirk3p}.
The only difference between the three- and four-point functions is that in
the case of the four-point function $\tilde{\Pi}$ is not free of the
middle variables $l_2$ and $k_2$. But this only results in a proliferation
of terms (more contributing residues)
compared with the three-point functions, while the structure
of the final integral representation remains unchanged.

It was the objective of this short note
to show that similar integral representations appear for the case of
three- and four-point functions.
In the three-point case, these integral representations
are the starting point for numerical approaches valid in arbitrary
kinematical regimes \cite{CKK}, and we hope that we can obtain similar
results for the box-functions in the future \cite{CK},
following the recipe as outlined here.\\[1cm]
{\Large Acknowledgements\\}
It is a pleasure to thank Andrzej Czarnecki for stimulating discussions
on the subject. This work was supported by the Australian Research
Council under grant number A69231484.


\begin{thebibliography}{99}
\bibitem{davydychev}
N.~I.~Ussyukina, A.~I.~Davydychev, Phys.~Lett.B305 (1993) 136.
\bibitem{CKK}
A.~Czarnecki, U.~Kilian, D.~Kreimer,
New representations of two-loop propagator and vertex functions,
MZ-TH/94-13.
\bibitem{CK}
A.~Czarnecki, D.~Kreimer, work in progress.
\bibitem{dirk3p}
D.~Kreimer, Phys.~Lett.B292 (1992) 341.
\bibitem{dirk2p}
D.~Kreimer, Phys.~Lett.B273 (1992) 277.
\bibitem{Itz}
C.~Itzykson, J.-B.~Zuber, Quantum Field Theory, Mc Graw-Hill (1980).
\bibitem{strocchi}
F.~Strocchi, General Properties of Quantum Field Theory,
World Scientific Lect.~Notes in Physics 51 (1993).
\bibitem{thiess}
F.~B.~Thiess, J.~Math.~Phys.9 (1968) 305.
\bibitem{wightman}
A.~S.~Wightman, Phys.~Rev.~101 (1956) 860.
\bibitem{dirkt}
D.~Kreimer, Mod.~Phys.~Lett.A9 (1994) 1105.
\end{thebibliography}
\end{document}